\documentclass[showpacs,preprintnumbers,amsmath,amssymb]{revtex4}
\usepackage{graphicx}
\usepackage{dcolumn}
\usepackage{bm}

\makeatletter
\def\simleq{\mathrel{\mathpalette\gl@align<}}
\def\simgeq{\mathrel{\mathpalette\gl@align>}}
\def\gl@align#1#2{\lower.6ex\vbox{\baselineskip\z@skip\lineskip\z@
     \ialign{$\m@th#1\hfill##\hfil$\crcr#2\crcr\sim\crcr}}}
\makeatother

\newcommand{\bra}{\langle}
\newcommand{\ket}{\rangle}
\newcommand{\braket}[1]{\bra #1 \ket}
\newcommand{\qq}{\braket{\bar{q}q}}

\newcommand{\qGq}{g\braket{\bar{q}\sigma_{\mu\nu}G_{\mu\nu} q}}

\begin{document}
\draft
\tighten

\title{The Quark-Gluon Mixed Condensate 
$g\langle\bar{q}\sigma_{\mu\nu}G_{\mu\nu} q\rangle$
in SU(3)$_c$ Quenched Lattice QCD}
\author{Takumi Doi$^1$
        \footnote{E-mail : doi@th.phys.titech.ac.jp},
Noriyoshi Ishii$^2$ ,
Makoto Oka$^1$ and
Hideo Suganuma$^1$ }
\affiliation{$^1$ Department of Physics, Tokyo Institute of Technology, 
\\
Oh-Okayama 2-12-1, Meguro, Tokyo 152-8551, Japan }
\affiliation{$^2$ Radiation Laboratory, 
The Institute of Physical and Chemical Research (RIKEN),
\\
Hirosawa 2-1, Wako, Saitama, 351-0198, Japan}



\begin{abstract}

Using the SU(3)$_c$ lattice QCD with the Kogut-Susskind fermion 
at the quenched level, 
we study the quark-gluon mixed condensate 
$g\langle\bar{q}\sigma_{\mu\nu}G_{\mu\nu} q\rangle$,
which is another chiral order parameter.
For each current quark mass, $m_q=21, 36$ and $52$ MeV, 
we generate 100 gauge configurations
in the $16^4$ lattice with $\beta = 6.0$, and perform the 
measurement of 
$g\langle\bar{q}\sigma_{\mu\nu}G_{\mu\nu} q\rangle$
at 16 points in each gauge configuration.
Using the 1600 data for each $m_q$, 
we find 
$m_0^2 \equiv 
g\langle\bar{q}\sigma_{\mu\nu}G_{\mu\nu} q\rangle / 
\langle\bar{q}q\rangle
\simeq 2.5$ GeV$^2$ 
at the lattice scale
in the chiral limit.
The large value of 
$g\langle\bar{q}\sigma_{\mu\nu}G_{\mu\nu} q\rangle$
suggests its importance in
the operator product expansion in QCD.

\end{abstract}



\maketitle


\section{Introduction}
\label{sec:intro}

The main signature of the non-perturbative nature of 
quantum chromodynamics (QCD) is its nontrivial vacuum structure,
which is represented by various condensates, or 
vacuum expectation values.
For instance, the quark condensate $\qq$ is a standard order parameter of 
spontaneous chiral symmetry breaking in QCD, 
and it determines properties of hadrons, especially the 
pion and the other pseudoscalar Nambu-Goldstone bosons.
In the gluonic sector, the gluon condensate $\langle G_{\mu\nu}G^{\mu\nu} \rangle$ is an important quantity 
associated with the trace anomaly in QCD, 
and the topological susceptibility $\langle Q^2 \rangle$ is responsible for the large $\eta'$ mass 
due to the U$_A$(1) anomaly.
Recently, the behavior of the various condensates at finite temperature/density is
a subject of intensive research in the context of the QCD phase diagram,
particularly the transition to the quark-gluon-plasma phase.

Among various condensates, we emphasize here the importance 
of the quark-gluon mixed condensate 
$\qGq \equiv 
{g\braket{\bar{q}\sigma_{\mu\nu}G_{\mu\nu}^A \frac{1}{2}\lambda^A q}}$.
First, the mixed condensate represents a direct correlation 
between quarks and gluons in the QCD vacuum. 
In this point, the mixed condensate differs from the above-mentioned condensates even at the qualitative level.
Second, this mixed condensate is another chiral order parameter 
of the second lowest dimension
and it flips the chirality of the quark as 
\begin{eqnarray}
\qGq = g\braket{\bar{q}_R\ (\sigma_{\mu\nu}G_{\mu\nu})\ q_L}
+ g\braket{\bar{q}_L\ (\sigma_{\mu\nu}G_{\mu\nu})\ q_R}.
\end{eqnarray}
Note here that the mixed condensate plays a relevant role in 
the operator product expansion (OPE) in QCD
as the next-to-leading chiral variant operator.

Also for the low-energy phenomena of hadrons,
the mixed condensate is found to be important 
through the framework of the QCD sum rule~\cite{SVZ},
which connects
the various condensates in OPE 
and the hadronic properties with the help of the dispersion relation.
The condensates are determined phenomenologically 
so as to reproduce 
various hadronic properties systematically, considering 
the Borel stability of the sum rules~\cite{RRY}.
For instance, in the standard QCD sum rule,
the nucleon mass $m_N$ and the delta mass $m_\Delta$ 
are given in terms of $\qq$ and $\qGq$ as\cite{Ioffe,Dosch}
\begin{eqnarray}
\lambda_N^2 m_N e^{-m_N^2/M^2} 
&=& \frac{1}{(2\pi )^2} M^4 (-\qq )+ 0 + {\cal O}(\mbox{7 dim. condensates}), \\
\lambda_\Delta^2 m_\Delta e^{-m_\Delta^2/M^2}
&=& \frac{4}{3(2\pi )^2} M^4 (-\qq ) - \frac{2}{3(2\pi )^2} M^2 (-\qGq ) + 
{\cal O}(\mbox{7 dim. condensates}) \\
&=& \frac{4}{3(2\pi )^2} M^4 \left( 1 - \frac{m_0^2}{2M^2} \right) (-\qq )
+ {\cal O}(\mbox{7 dim. condensates}),
\end{eqnarray}
where 
$M$ denotes the Borel mass and $\lambda_N$ and $\lambda_\Delta$ 
parameters in the QCD sum rule~\cite{Ioffe,Dosch}.
Here, we have used the standard parameterization as
\begin{eqnarray}
m_0^2 \equiv \qGq / \qq.
\end{eqnarray}
In the QCD sum rules,
the value $m_0^2 \simeq 0.8\pm 0.2\ {\rm GeV}^2$ has been 
proposed as a result of the 
phenomenological analyses~\cite{Bel,RRY2,Ovchi,Narison1}.
Therefore, for the sum rule on $\Delta$,
the second OPE term,
which is proportional
to the mixed condensate, amounts to the same magnitude 
to the leading OPE term, if we take the Borel mass $M$ equals 
to the typical baryon mass as $1$GeV.
From these equations, one finds that the condensate $\qGq$ 
has large effects on the $N$-$\Delta$ splitting~\cite{Dosch}.

The condensate $\qGq$ is also important in the light-heavy 
meson systems~\cite{Dosch2}, 
since  the term $m_H \cdot \qGq$ proportional to the heavy quark mass $m_H$ 
contributes significantly in OPE of the corresponding sum rule.
Furthermore, through the direct mixing of $q$, $\bar q$ and gluons, 
the mixed condensate $\qGq$ directly contributes
in the exotic meson systems~\cite{Latorre}.

Needless to say, it is desirable to estimate $\qGq$,
not only by the phenomenological parameter fitting in QCD sum rules, 
but also by a direct calculation from QCD. 
For this purpose, the lattice QCD Monte Carlo simulation~\cite{Wilson}
is a powerful tool.
With this method, the condensates can be directly calculated
from QCD, keeping the non-perturbative effect.
However, in spite of the importance of $\qGq$, 
there was only one preliminary lattice QCD study
for $\qGq$ done about 15 years ago~\cite{K&S}.
This pioneering study~\cite{K&S} gave an estimate
$m_0^2 \sim 1.1 {\rm GeV}^2$, but this result is not conclusive yet
because the simulation was done with insufficient statistics 
using a small and coarse lattice: 
the authors used only 5 gauge configurations on the $8^4$ lattice with $\beta=5.7$, 
and calculated the condensates at only 1 space-time point for each gauge configuration.

Therefore, in this paper, we present the calculation of $\qGq$ in 
lattice QCD with a large and fine lattice and with high statistics.
We perform the measurement of $\qGq$ as well as $\qq$
in the SU(3)$_c$ lattice at the quenched level. 
Since these condensates are chiral order parameters,
we adopt the Kogut-Susskind (KS) fermion,
which keeps the explicit chiral symmetry in the massless quark limit.
We generate 100 gauge configurations and pick up 16 space-time points 
for each configuration to calculate the condensates.
Therefore, we obtain 1600 data for each quark mass and each $\beta$.
We perform reliable estimate of the condensates
with this high statistics.

This paper is organized as follows.
In Sec.~\ref{sec:formalism},
we explain our formalism to calculate
the condensates.
In Sec.~\ref{sec:results&check},
we present the lattice QCD data,
and discuss its reliability
by performing several checks.
Sec.~\ref{sec:summary} is devoted to summary and concluding remarks.



\section{Formalism}
\label{sec:formalism}

In  this section,  we  describe  the formalism  
on the calculation of 
the condensates   $\qq$   and  $\qGq$
in SU(3)$_c$  quenched lattice QCD.
Note that, 
even without the dynamical quark effects, 
the quenched lattice QCD calculations
have reproduced various hadronic properties in good agreement 
with empirical values~\cite{Rothe}.
Moreover, the characteristics of the quenched simulation
are well under control owing to the accumulated knowledge.
Therefore, it is worth performing the quenched lattice QCD calculation 
before proceeding to the full QCD calculation 
as a next step.

The lattice QCD is formulated in terms of the link-variable $U_\mu(s) \equiv \exp[-iagA_\mu(s)]$ 
on the lattice with spacing $a$, instead of the continuum gluon field $A_\mu(x)$.
For the gauge sector, we adopt the standard Wilson action as
\begin{eqnarray}
S_G = \sum_s \sum_{\mu > \nu} \beta 
\left[\ 1-\frac{1}{N_c}{\rm Re\ Tr\ }U_{\mu\nu}(s) \right],
\end{eqnarray}
with $\beta \equiv 2N_c/g^2$ and 
the plaquette operator $U_{\mu\nu}(s)$  on the $(\mu,\nu)$-plane,
which is described by
\begin{eqnarray}
U_{\mu\nu}(s)\equiv U_\mu(s)U_\nu(s+\mu)U^\dagger_\mu(s+\nu)U^\dagger_\nu(s). 
\end{eqnarray}
For the fermion action, we adopt the KS-fermion.
As the advantage of the KS-fermion,
its action takes a simple form 
and preserves the explicit chiral symmetry in massless quark limit, $m=0$.
The latter property of the KS-fermion is desirable for our study,
since both of the  condensates  $\qq$  and  $\qGq$
are   expected  to   be sensitive to explicit chiral  symmetry breaking
as chiral order parameters.
%

We comment here on the other lattice fermions briefly.
The domain-wall fermion would be attractive
from the viewpoint of chiral symmetry. However,
its simulation is much more expensive 
in comparison with the KS-fermion.
In addition,
there are ambiguities originating from the newly introduced simulation 
parameters such as the domain-wall height.
The Wilson and  the clover fermions would not  be appropriate for our
purpose, because  they have a serious disadvantage  from the viewpoint
of chiral symmetry.
Specifically, the action for these fermions contains the term
\begin{eqnarray}
{\cal L}_{\rm E \chi SB} \propto \ \ 
\bar{q}(s) \left[\ U_\mu(s) q(s+\mu) + U^\dag_\mu(s-\mu) q(s-\mu) 
		   - 2 q(s) 
           \right],
\end{eqnarray}
which explicitly breaks chiral symmetry even for $m=0$.
Although this term  vanishes in the continuum limit,  the chiral order
parameters inevitably  suffer the nontrivial contamination
from   this  unphysical   term   at  finite   lattice  spacing.   This
uncontrollable contamination should be avoided.

The action for the KS-fermion~\cite{Rothe} with the mass $m$ is described by 
\begin{eqnarray}
S_F = \frac{1}{2} \sum_{s,\mu} \eta_\mu (s) \bar{\chi}(s)
 \left[ U_\mu (s) \chi (s+\mu ) - U^\dag_\mu (s-\mu ) \chi (s-\mu ) 
 \right]
 + ma\sum_s \bar{\chi}(s) \chi (s),
\label{eq:ks-action}
\end{eqnarray}
where  $\bar{\chi}$ and $\chi$ are Grassmann fields which have no spinor degrees of freedom, and
$\eta_\mu (s)$ is the staggered phase defined as $\eta_\mu (s) \equiv (-1)^{s_1+\cdots +s_{\mu -1}}$, {i.e.},
\begin{eqnarray}
\eta_1(s)=1,\ \
\eta_2=(-1)^{s_1},\ \
\eta_3=(-1)^{s_1+s_2},\ \
\eta_4=(-1)^{s_1+s_2+s_3}.
\end{eqnarray}
In order to make the definition of the sign of $\qGq$ unambiguous,
we note here that the definition of the continuum covariant derivative 
is $ D_\mu \equiv \partial_\mu - i g A_\mu $, 
corresponding to the definition of $U_\mu \equiv e^{-iagA_\mu}$.

In this formalism, the quark field $q$ is introduced as an ${\rm SU}(N_f=4)$ spinor field.
The explicit  relation between  the quark
field $q$ and the spinless Grassmann field $\chi$ is understood in the
following way.
When the gauge field is set to be zero, the quark field $q$ is expressed by 
\begin{eqnarray}
\label{eq:q-ks_trans}
q_i^f (x) 
&=&  \frac{1}{8}
\sum_{\rho}\
( \Gamma_\rho )_{if}\ 
\chi (x+\rho ), \\
\label{eq:Gamma}
\Gamma_\rho &\equiv& \gamma_1^{\rho_1} \gamma_2^{\rho_2} \gamma_3^{\rho_3} \gamma_4^{\rho_4},
\end{eqnarray}
where $\rho= (\rho_1,\rho_2,\rho_3,\rho_4)$ with $\rho_\mu \in \{0,1\}$ runs over the 16 sites 
in the $2^4$ hypercube. The indices $i$ and $f$ denote 
the spinor and the flavor indices, respectively.
When the gluon field is turned on, 
we insert
additional link-variables in Eq.~(\ref{eq:q-ks_trans})
in order to respect the gauge covariance.

The evaluation of the condensates amounts to the following expressions
as
\begin{eqnarray}
a^3 \qq 
&=& - \frac{1}{4}\sum_f {\rm Tr}\left[ S^f(x,x) \right],
\label{eq:condensates-qq-def} \\
a^5 \qGq
&=& - \frac{1}{4}\sum_f \sum_{\mu,\nu}{\rm Tr}
	\left[ S^f(x,x) \sigma_{\mu\nu} G_{\mu\nu}\right],
\label{eq:condensates-qGq-def} 
\end{eqnarray}
where the  trace ``${\rm  Tr}$'' is taken with respect to both
the spinor and  the color
indices,  and $S^f(y,x)$  denotes the  Euclidean quark  propagator for
$f$-th flavor as
\begin{eqnarray}
S^f(y,x) = \braket{q^f(y) \bar{q}^f(x)}. 
\end{eqnarray}

\begin{figure}
\caption{\label{fig:qq-diag}
The diagrammatic representation of $\qq$ in terms of the
spinless Grassmann field $\chi$
on the lattice. The solid curve with the arrow denotes the propagation
of $\chi$.
}
\begin{center}
\includegraphics[scale=0.6]{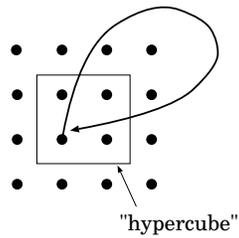}
\end{center}
\end{figure}

In terms of $\chi$ and $\bar{\chi}$-fields,
the flavor-averaged quark condensate is rewritten on the lattice as
\begin{eqnarray}
a^3 \qq 
&=& - \frac{1}{2^8}\sum_\rho
       {\rm Tr}\left[ \Gamma_\rho \Gamma_\rho^\dag\ \braket{\chi(x+\rho ) \bar{\chi}(x+\rho )} \right].
\label{eq:qq-def1}
\end{eqnarray}
The corresponding diagram is shown in figure~\ref{fig:qq-diag}.

On the other hand, the flavor-averaged quark-gluon mixed condensate is given by
\begin{eqnarray}
\label{eq:qGq-def1}
a^5 \qGq  
&=& - \frac{1}{2^8} \sum_{\mu,\nu} \sum_\rho
{\rm Tr}\left[\
  {\cal U}_{\pm\mu,\pm\nu}(x+\rho)\ 
\Gamma_{\rho'} \Gamma_{\rho}^\dag\
\braket{\chi (x+\rho') 
\bar{\chi}(x+\rho)} \
\sigma_{\mu\nu}\ 
  G_{\mu\nu}^{\rm lat}(x+\rho)\ 
\right], \\
\label{eq:rho-2Ddiag}
&& \rho' \equiv \rho \pm \mu \pm \nu, \nonumber
\end{eqnarray}
where the sign $\pm$ is taken such that the sink point $(x+\rho') = (x+\rho\pm\mu\pm\nu)$
belongs to the same hypercube of the source point $(x+\rho)$.
Here, in order to respect the gauge covariance,
we have used in Eq.~(\ref{eq:qGq-def1})
%
%
\begin{eqnarray}
{\cal U}_{\pm\mu,\pm\nu}(x) \equiv 
	\frac{1}{2}\left[\ U_{\pm\mu} (x) U_{\pm\nu} (x\pm\mu ) 
		+ U_{\pm\nu} (x) U_{\pm\mu} (x\pm\nu )\ \right],
\label{eq:qGq-def-U}
\end{eqnarray}
where we use the definition of $U_{-\mu} (x) \equiv U^\dag_{\mu}(x-\mu)$.

\begin{figure}
\caption{\label{fig:qGq-diag}
The diagrammatic representation of the two ingredients in
$\qGq$ in terms of the spinless Grassmann field $\chi$
and the gluon field on the lattice.
(a) The left diagram shows the propagation of $\chi$, with the 
insertion of gauge links.
The solid curve with the arrow denotes the propagation of $\chi$,
and the curly lines denote the inserted gauge 
link ${\cal U}_{\mu,\nu}$.
(b) The right diagram shows the gluon field strength 
$G_{\mu\nu}^{\rm lat}$, where each loop
of the curly line denotes a plaquette operator.
}
\begin{center}
\includegraphics[scale=0.6]{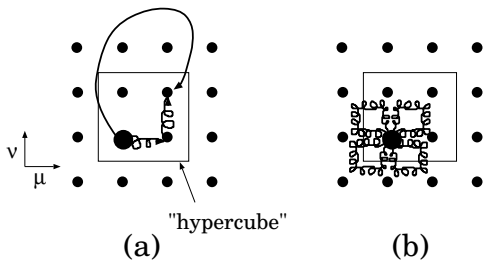}
\end{center}
\end{figure}

On the gluon field strength $G_{\mu\nu}$,
we adopt the clover-type definition on the lattice,
\begin{eqnarray}
G_{\mu\nu}^{\rm lat}(s) = \frac{i}{16} \sum_A \lambda^A \ {\rm Tr}
\left[ 
 \lambda^A\{U_{\mu\nu}(s) +U_{\nu\,-\!\mu}(s)+U_{-\!\mu\,-\!\nu}(s)+U_{-\!\nu\,\mu}(s) \}
-\lambda^A \{\mu \leftrightarrow \nu \} 
\right],
\label{eq:clover}
\end{eqnarray}
where $\lambda^A$ ($A=1,2,\cdots,8$)  denotes the color SU(3) Gell-Mann
matrix normalized as $\mbox{Tr}(\lambda^A\lambda^B) =
2 \delta^{AB}$.
In figure~\ref{fig:qGq-diag},
we show the diagrams corresponding to 
Eqs.~(\ref{eq:qGq-def1}) and (\ref{eq:clover}).

In the continuum limit, 
Eq.~(\ref{eq:clover}) leads to 
\begin{eqnarray}
G_{\mu\nu}^{\rm lat}(s) 
\rightarrow 
a^2 \left[\ g G^A_{\mu\nu}(s) \frac{\lambda^A}{2} + {\cal O}(a^2)\ \right].
\end{eqnarray}
It is worth mentioning that this definition has no ${\cal O}(a)$ discretization error. 
On the other hand, in Ref.\cite{K&S}, the authors adopted a
simple insertion of the gluon field strength,
\begin{eqnarray}
a^5 \qGq &=& \frac{1}{4} \sum_f
\bar{q}^f(s) \sigma_{\mu\nu} U_\mu(s) U_\nu(s+\mu) q^f(s+\mu+\nu),
\label{eq:qGq-def_K&S}
\end{eqnarray} 
which contains ${\cal O}(a)$ error.
Although both of the definitions, Eqs.~(\ref{eq:qGq-def1}) and (\ref{eq:qGq-def_K&S}),
coincide to the $\qGq$ in the continuum limit,
our definition of $G_{\mu\nu}$ 
will give less systematic errors
in the actual lattice simulations with finite lattice spacing $a$.



\section{The lattice QCD results}
\label{sec:results&check}

\subsection{Lattice QCD results for $\qGq$}
\label{sec:results}

We calculate the condensates $\qq$ and $\qGq$ using the SU(3)$_c$ lattice QCD at the quenched level.
The Monte Carlo simulation is performed with the standard Wilson action for $\beta=5.7, 5.8$ and $6.0$ 
on the $8^4, 12^4$ and $16^4$ lattice, respectively.
The pseudo-heat-bath algorithm is adopted for the update of the gauge configuration.
After 1000 sweeps for the thermalization, 
we pick up 100 gauge configurations for every 500 sweeps.
The lattice unit $a$ is determined so as to reproduce 
the string tension $\sigma = 0.89 {\rm GeV/fm}$~\cite{rabbit:3Q}.
In Table~\ref{tab:gauge}, we summarize the lattice parameters for the gauge configuration.
We note that the physical volume $V$ is roughly the same for the three calculations with different $\beta$.

We use the quark mass parameter, $m = 21, 36$ and $52$ MeV, which correspond to 
$ma = 0.0105,\ 0.0184$\ and\ $0.0263$ for $\beta = 6.0$, respectively.
Also for $\beta=5.7$ and $5.8$, we use the same values of the physical quark mass $m$. 
The corresponding values of $ma$ are also tabulated in Table~\ref{tab:gauge}.


\begin{table}
\begin{center}
\caption{\label{tab:gauge}
The lattice QCD parameter $\beta\equiv 2N_c/g^2$ 
and the lattice size used in the simulation.
The lattice spacing $a$, the physical volume $V$ and 
the adopted values of the current quark mass $ma$ are also 
listed for each $\beta$.
As for the current quark mass, the corresponding physical values
are $m=21, 36$ and $52$ MeV from the left.
}
\vspace{0.5cm}
\begin{tabular}
{cccccccc}
\hline
\hline
 	  $\beta$ & lattice size & $a$ [fm] &  $V$[fm$^4$]  & & $ma$ & 
\\ 
\hline
5.7  &  $8^4$    & 0.19        &   $(1.5)^4$   & 0.0200 & 0.0350 & 0.0500  \\
5.8  & $12^4$    & 0.14        &   $(1.7)^4$   & 0.0147 & 0.0258 & 0.0368  \\
6.0  & $16^4$    & 0.10        &   $(1.6)^4$   & 0.0105 & 0.0184 & 0.0263  \\ 
\hline
\hline
\end{tabular}
\end{center}
\end{table}


In the determination of the Euclidean propagator 
$\braket{\chi (y_\chi)\bar{\chi}(x_\chi)}$,
we solve the matrix inverse
equations iteratively
using the CG, BiCGSTAB and MR algorithms,
until the residual error $r^2$ 
becomes small enough to satisfy $r^2  < 10^{-8} (\beta = 6.0)$ or $r^2 < 10^{-10} (\beta = 5.7, 5.8)$.
By checking the differences of the results among these algorithms,
we confirm the numerical errors are smaller than
the statistical errors for all $\beta$.
For the Grassmann $\chi$-field, the anti-periodic condition is imposed.
The dependence on the boundary condition will be discussed later.

In the KS-fermion formalism, the source point $x_\chi \equiv x+\rho$ of the
$\chi$-field is taken to be on the hypercubic site around
the physical source point $x$.
%
We take 16 physical space-time source points $x$ in each gauge
configuration as follows:
on the lattices with the volume $(2n)^4=8^4,12^4$ and $16^4$, 
we take $x=(x_1,x_2,x_3,x_4)$ 
with $x_\mu \in \{0, n\}$ 
in the lattice unit.
For each physical space-time point $x$, 
we take the sum over $\rho$ in the hypercube, 
corresponding to
the flavor and spinor contractions.
%
%
%
For each $\beta$ and $m$, we calculate the flavor-averaged condensates 
according to Eqs.~(\ref{eq:qq-def1}) and (\ref{eq:qGq-def1}),
and average them over the 16 physical space-time points and 
100 gauge configurations.
Statistical errors are calculated using the jackknife error estimate.

\begin{figure}
\caption{\label{fig:plot}
The bare condensates $a^3\qq$ and $a^5\qGq$ plotted against the quark mass $ma$ 
at $\beta=6.0$.
The dashed lines denote the best linear extrapolations, 
and the cross symbols correspond to the values in the chiral limit. 
The jackknife errors are hidden in the dots.
}
\vspace*{8mm}
\begin{center}
\includegraphics[scale=0.75]{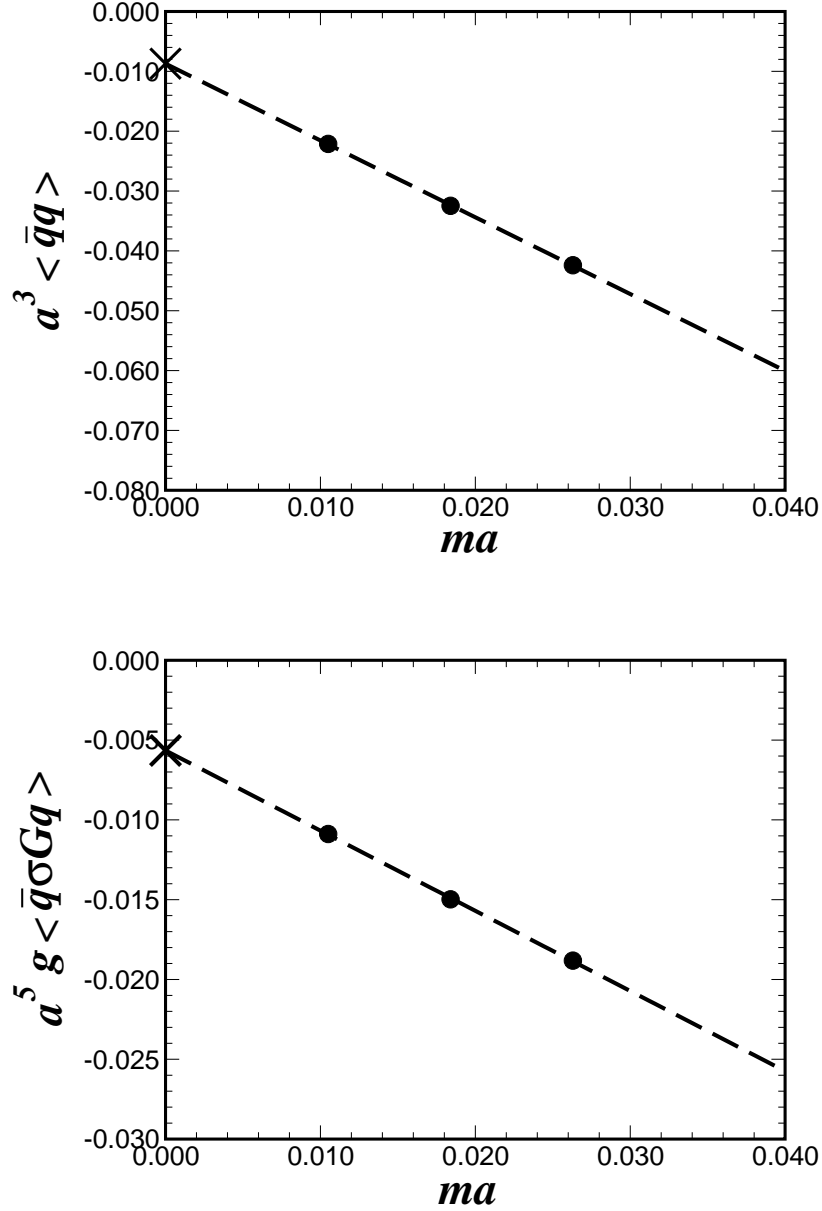}
\end{center}
\end{figure}

Figure~\ref{fig:plot} shows the 
values of the bare condensates $a^3\qq$ and $a^5\qGq$ against the quark mass $ma$.
We emphasize that the jackknife errors are almost negligible, 
due to the high statistics of $1600$ data for each quark mass.
From figure~\ref{fig:plot}, both $\qq$ and $\qGq$ show 
a clear linear behavior against the quark mass $m$.
This feature is also found for $\beta=5.7$ and $5.8$.
Therefore, we fit the data  with a linear function and determine the condensates in the chiral limit.
The results are summarized in Tables
~\ref{tab:mass-beta-6.0},~\ref{tab:mass-beta-5.8} and ~\ref{tab:mass-beta-5.7}.

\begin{table}
\begin{center}
\caption{\label{tab:mass-beta-6.0}
The numerical results of $a^3\qq$ and $a^5\qGq$ for various $ma$ 
in SU(3)$_c$ lattice QCD with $\beta$=6.0 and $16^4$.
The last column denotes their values in the chiral limit obtained by the linear chiral extrapolation.
}
\begin{tabular}{ccccc}
\hline
\hline
	   &  $ma=0.0263$     &  $ma=0.0184$     &  $ma=0.0105$      &  chiral limit\\
\hline
$a^3\qq$   &  $-0.04240(16)$  &  $-0.03247(15)$ &  $-0.02212(16)$  & 
$-0.00872(17)$ \\
$a^5\qGq$  &  $-0.01882(15)$  &  $-0.01498(14)$ &  $-0.01088(14)$  & 
$-0.00565(14)$ \\
\hline
\hline
\end{tabular}
\end{center}
\end{table}

\begin{table}
\begin{center}
\caption{\label{tab:mass-beta-5.8}
The numerical results of $a^3\qq$ and $a^5\qGq$ for various $ma$ 
in SU(3)$_c$ lattice QCD with $\beta$=5.8 and $12^4$.
The last column denotes their values in the chiral limit obtained by the linear chiral extrapolation.
}
\begin{tabular}{ccccc}
\hline
\hline
	   &  $ma=0.0368$     &  $ma=0.0258$     &  $ma=0.0147$      &  chiral limit\\
\hline
$a^3\qq$   &  $-0.08031(35)$  &  $-0.06667(35)$  &  $-0.05159(36)$  & 
$-0.03271(38)$ \\
$a^5\qGq$  &  $-0.03883(32)$  &  $-0.03375(32)$  &  $-0.02783(31)$  & 
$-0.02064(32)$ \\
\hline
\hline
\end{tabular}
\end{center}
\end{table}

\begin{table}
\begin{center}
\caption{\label{tab:mass-beta-5.7}
The numerical results of $a^3\qq$ and $a^5\qGq$ for various $ma$ 
in SU(3)$_c$ lattice QCD with $\beta$=5.7 and $8^4$.
The last column denotes their values in the chiral limit obtained by the linear chiral extrapolation.
}
\begin{tabular}{ccccc}
\hline
\hline
	   &  $ma=0.050$      &  $ma=0.035$      &  $ma=0.020$       &  chiral limit\\
\hline
$a^3\qq$   &  $-0.12346(68)$  &  $-0.10788(73)$  &  $-0.09017(80)$  & 
$-0.06833(90)$ \\
$a^5\qGq$  &  $-0.06200(53)$  &  $-0.05695(56)$  &  $-0.05064(60)$  & 
$-0.04327(66)$ \\
\hline
\hline
\end{tabular}
\end{center}
\end{table}


\subsection{Check on the systematic uncertainty}
\label{sec:check}

In this section, we check the reliability of our lattice QCD results.
We first consider the finite volume artifact.
As indicated by the Banks-Casher formula~\cite{Banks},
\begin{eqnarray}
\lim_{m\rightarrow 0} \lim_{V\rightarrow \infty} \qq 
&=& -\pi \frac{\rho(0)}{V}, 
\end{eqnarray}
for the spectral density $\rho(\lambda)$ of the Dirac operator,
the total volume $V$ should be large enough before the quark mass goes to zero.
In order to estimate this finite volume artifact, 
we carry out the same calculation imposing the periodic boundary condition 
on the Grassmann fields $\chi$ and $\bar{\chi}$,
instead of the anti-periodic boundary condition,
keeping the other parameters same.
If the lattice total volume is too small, the quark
propagates over the total volume, and
the results would be sensitive to the boundary conditions.
Thus, 
the difference will indicate ambiguity from the finite volume artifact.

\begin{table}
\begin{center}
\caption{\label{tab:mass-beta-6.0-periodic}
The lattice results of $a^3\qq$ and $a^5\qGq$ with $\beta$=6.0 and $16^4$ 
in the case of the periodic boundary condition for the fermion field. 
This calculation is done for the check of the boundary effect and the 
finite volume artifact.
}
\begin{tabular}{ccccc}
\hline
\hline
	   &  $ma=0.0263$     &  $ma=0.0184$     &  $ma=0.0105$      &  chiral limit\\
\hline
$a^3\qq$   &  $-0.04236(16)$  &  $-0.03240(16)$  &  $-0.02200(16)$   & 
$-0.00854(17)$ \\
$a^5\qGq$  &  $-0.01880(15)$  &  $-0.01493(14)$  &  $-0.01078(14)$   & 
$-0.00551(14)$ \\
\hline
\hline
\end{tabular}
\end{center}
\end{table}

We show in Table~\ref{tab:mass-beta-6.0-periodic} the
lattice results with the periodic boundary condition for $\beta=6.0$.
Comparing with Table~\ref{tab:mass-beta-6.0}, 
one finds that the difference is only 1\% level.
The similar results are obtained for $\beta=5.7$ and $5.8$.
Therefore, we conclude that the physical volume
$V \sim (1.6\ {\rm fm})^4$ in our simulations is large enough to 
avoid the finite volume artifact.

We next consider the discretization error.
As an advanced feature of the KS-fermion, 
the discretization 
error begins from ${\cal O}(a^2)$ on the lattice spacing $a$. 
This is because ${\cal O}(a)$ errors cancel with each other 
when the average over the SU(4) flavor is taken.
Therefore, 
there is no ${\cal O}(a)$ error originating from
the quark propagator in 
Eqs.~(\ref{eq:condensates-qq-def}) and (\ref{eq:condensates-qGq-def}).
On the other hand, there is an ambiguity 
coming from a particular choice
of the gauge link ${\cal U}_{\mu,\nu}$ 
in Eq.~(\ref{eq:qGq-def-U}),
which is introduced to respect the gauge covariance.
This ambiguity can be checked 
by changing the definition of $\qGq$, 
adopting a different path which connects 
the source point $(x+\rho)$ and the sink point $(x+\rho')$ 
in Eq.~(\ref{eq:qGq-def1}).
Specifically, instead of ${\cal U}_{\mu,\nu}$ in Eq.~(\ref{eq:qGq-def-U}), 
we examine the other product $\tilde{\cal U}_{\mu,\nu}$ as 
\begin{eqnarray}
\tilde{\cal U}_{\pm\mu,\pm\nu}(x) \equiv U_{\pm\mu} (x) U_{\pm\nu} (x\pm\mu ), 
\label{eq:qGq-def2-U}
\end{eqnarray}
and thus define $\qGq$ by
%
%
\begin{eqnarray}
a^5 \qGq  
&=& - \frac{1}{2^7} \sum_{\mu >\nu} \sum_\rho
{\rm Tr}\left[\
  \tilde{\cal U}_{\pm\mu,\pm\nu}(x+\rho)\ 
\Gamma_{\rho'} \Gamma_{\rho}^\dag\
\braket{\chi (x+\rho') 
\bar{\chi}(x+\rho)} \
\sigma_{\mu\nu}\ 
  G_{\mu\nu}^{\rm lat}(x+\rho)\ 
\right],
\label{eq:qGq-def2}
\end{eqnarray}
where $\rho' =  \rho \pm \mu \pm \nu$ and the sign $\pm$ is taken 
as before.
We perform the calculation for $\beta = 5.7$ and $ma = 0.050$. 
In this case, the lattice spacing $a$ is largest in our simulations, 
and therefore 
the discretization error is expected to be larger than the other cases 
for $\beta=5.8$ and $6.0$.
At each gauge configuration, we check the difference of the results
between the choice of Eq.~(\ref{eq:qGq-def-U})
and Eq.~(\ref{eq:qGq-def2-U}).
Since the typical difference is about 1\%, we conclude 
that the  discretization error is small enough,
which confirms the reliability of our lattice results.


\subsection{Determination of $m_0^2=\qGq/\qq$}
\label{sec:M0}

The values  of the  condensates in the  continuum limit  are to be obtained
after the renormalization.
To  this end,  the lattice  perturbation theory  has often been used,
although   it   is   afflicted with  uncertainty  originating   from   the
non-perturbative   effect.    In   principle,   the   non-perturbative
renormalization scheme  is desirable, which, however, requires  a lot of
computational power~\cite{Martinelli}.
Therefore we seek  for another way which can  reduce this uncertainty.
Here  we  provide   the  ratio  $m_0^2  \equiv  \qGq   /  \qq$,  where
some of the uncertainties are canceled with each other.
In particular, this ratio  is free  from the  uncertainty from  the wave
function  renormalization  of the  quark.
%
As  a  consequence,  the   results  become  more  reliable  with  less
uncertainties.
In  addition, the  dependence of  $m_0^2$  on the  lattice spacing  is
weakened
to $a^2$,  while  $\qq$ and $\qGq$  are  proportional to  $a^3$ and
$a^5$, respectively.
We note  that $m_0^2$  itself has the  following physical  meaning.
In the QCD sum rule,  $\qGq$ generally appears 
as the chiral variant term next to $\qq$ in  OPE.  
Therefore, $m_0^2$ is usually used 
without referring to the absolute  value of $\qGq$ itself,
and thus it represents the level of importance of $\qGq$ 
in OPE.

Now, we present the result of  the ratio $m_0^2$ using the bare results
in SU(3)$_c$ lattice QCD.
We adopt the results at  $\beta=6.0$, since its lattice spacing is the
finest in our calculations.
%
%
%
%
%
We find in the chiral limit
\begin{eqnarray}
m_0^2 \equiv \qGq / \qq  \simeq  2.5\ {\rm GeV}^2 \ (\beta = 6.0),
\end{eqnarray}
at the lattice cutoff scale as $a^{-1} \simeq 2$ GeV.
This large value of $m_0^2$ suggests the significance
 of the   mixed  condensate   in  OPE.
Although we  do not include  renormalization effect, this  result
itself is determined very precisely.



\section{Summary and Discussions}
\label{sec:summary}

We  have studied  the quark-gluon  mixed condensate  $\qGq$  using 
SU(3)$_c$ lattice QCD with  the Kogut-Susskind fermion at the quenched
level.
First,  we have  emphasized  that the  mixed  condensate is  one  of the  key
quantities in various quark hadron physics, especially in the baryon sector
such as the $N$-$\Delta$ splitting.
In spite of  its importance, the lattice QCD  studies of this quantity
have been limited to only  one preliminary study for 15
years.
Recently, due to the progress in lattice QCD Monte Carlo calculations,
it becomes possible to calculate  
this mixed condensate with   much better statistics
on a finer and larger lattice.
For each  quark mass of  $m_q=21, 36$ and $52$  MeV, we have  generated 100
gauge configurations  on the $16^4,  12^4$ and $8^4$ lattice with  $\beta =
6.0, 5.8$ and $5.7$, respectively.
We have  performed the measurements of  $\qGq$ as well as  $\qq$ at 16
physical
space-time points in each gauge configuration.
Using the 1600 data for each $m_q$,  we have found $m_0^2 \equiv \qGq / \qq
\simeq 2.5$ GeV$^2$ in the chiral limit
at the lattice scale $a^{-1} \simeq 2$ GeV corresponding to $\beta=6.0$.
We have checked that 
the   systematic  and   statistical   errors  are   almost negligible.
Therefore, the value of $m_0^2$ at the lattice scale has been well
determined in this calculation.
%


Finally, we compare our result with  the standard value employed in the QCD sum
rule.
%
To this end,  we change the renormalization point
from  $\mu\simeq\pi/a$ to $\mu\simeq 1$
GeV corresponding to the QCD sum rule.
Following  Ref.~\cite{K&S},  we  first  take  the lattice results  of  the
condensates  as  the starting  point  of  the  flow, then  rescale  the
condensates using the anomalous dimensions evaluated 
in a perturbative manner.
We have the following rescaled condensates as~\cite{Narison2},
%
%
%
%
%
%
\begin{eqnarray}
\label{eq:qq-anomalous}
&  \qq \Big|_{\mu} & = 
\left( \frac{\alpha_s(\mu)}{\alpha_s(\pi/a)} \right)^{-4/b_0} \ \ 
\qq \big|_{\pi/a}, \\
\label{eq:qGq-anomalous}
& \qGq \Big|_{\mu} & = 
\left( \frac{\alpha_s(\mu)}{\alpha_s(\pi/a)} \right)^{2/(3b_0)} \ \ 
\qGq \Big|_{\pi/a},
\end{eqnarray}
where we use the one-loop formula,
$\alpha_s(\mu) = \frac{4\pi}{b_0 \ln (\mu^2 / \Lambda_{\rm QCD}^2 )}$
with $\Lambda_{\rm QCD} = 200-300 \mbox{MeV}$ and 
$b_0 = (11/3) N_c - (2/3) N_f$.
(The anomalous dimension given in Refs.~\cite{Beneke,Alad}
for $\qGq$ is different from Ref.~\cite{Narison2}.
However, this difference does not change our
semi-quantitative analysis here.)
%
%
For the  case of the quenched lattice  QCD, we adopt  $N_f=0$.  By using
our bare  lattice QCD  results at $\beta=6.0$,  we obtain
$ m_0^2  \big|_{\mu} \equiv  \qGq/\qq  \big|_{\mu} \sim  3.5-3.7 \  
{\rm GeV}^2$
at $\mu=1$GeV, 
from $\qq  \big|_{\mu} \sim -  (0.0477 - 0.0506 )  \ {\rm GeV}^3 
=  - (0.36-0.37  {\rm GeV})^3$ and 
$\qGq \big|_{\mu}  \sim - (0.176- 0.177)  \ {\rm GeV}^5$.
Comparing with the standard value of  $m_0^2 = 0.8 \pm 0.2$ GeV$^2$ in the
QCD sum rule,  our calculation results in a  rather large value. Note
that  the instanton  model has  made  a slightly  larger estimate  as
$m_0^2 \simeq 1.4$ GeV$^2$ at $\mu\simeq 0.6$ GeV~\cite{Polyakov}.
For a more definite determination of $m_0^2$, 
the  renormalization procedure should
be performed  more carefully,  which is also  expected to  improve the
value  of $\qq$  simultaneously.  In  principle,  the non-perturbative
renormalization  scheme is    most desirable,  which would, however,
require a significant calculation cost \cite{Martinelli}.

We  again  emphasize  that  the  mixed condensate  $\qGq$  plays  very
important roles  in various contexts in quark  hadron physics.  
Hence, it is preferable to perform  further studies.  
In particular, the dynamical  quark   effects  
would be nontrivial, since
the mixed condensate includes quark field.
The  thermal  effects  are also interesting in relation to chiral 
restoration, because the  mixed  condensate is  another chiral  order
parameter.  Actually, we are in progress with these two studies on
the lattice \cite{DOIS:T}.
%
%
Considering the RHIC  project, it  becomes  more  and  more important  to
understand the nature of finite temperature QCD.
%
%
Therefore, it is quite desirable to determine thermal effects
on the condensates with lattice QCD in understanding the finite 
temperature QCD.
%


\acknowledgments
We would like to thank Dr. H. Matsufuru for his useful comments on 
the programming technique. 
This work is supported in part by the Grant for Scientific Research 
(No.11640261, No.12640274 and No.13011533) 
from the Ministry of Education, Culture, Science and Technology, Japan.
T.D. acknowledges the support by the JSPS
(Japan Society for the Promotion of Science) Research Fellowships 
for Young Scientists.
The Monte Carlo simulations have been performed on the 
NEC SX-5 supercomputer at Osaka University.







\begin{references}

\bibitem{SVZ}       {M.A. Shifman, A.I. Vainshtein and V.I. Zakharov,
                            Nucl. Phys. {\bf B 147}, 385 (1979), {\it ibid.}, 448 (1979).}
%

\bibitem{RRY}       { L.J. Reinders, H. Rubinstein and S. Yazaki, 
                           Phys. Rep. {\bf 127}, 1 (1985) 
			and references therein.}

\bibitem{Ioffe}     {B. L. Ioffe,
                            Nucl. Phys.  {\bf B188}, 317 (1981),
			    {\it Erratum-ibid.} {\bf B191}, 591 (1981).}

\bibitem{Dosch}     {H.G. Dosch, M. Jamin, and S. Narison,
                        Phys. Lett. {\bf B220}, 251 (1989). }

\bibitem{Bel}       {V.M. Belyaev and B.L. Ioffe,
			Sov. Phys. JETP {\bf 56}, 493 (1982).}

\bibitem{RRY2}     {L.J. Reinders, H.R. Rubinstein, and S. Yazaki,
                        Phys. Lett. {\bf B120}, 209 (1983).}

\bibitem{Ovchi}     {A.A. Ovchinnikov and A.A. Pivovarov,
                        Sov. J. Nucl. Phys. {\bf 48}, 721 (1988) 
                        (Yad. Fiz. {\bf 48}, 1135 (1988)).}

\bibitem{Narison1}     {S. Narison,
                        Phys. Lett. {\bf B210}, 238 (1988).}

\bibitem{Dosch2}     {H.G. Dosch, and S. Narison,
                        Phys. Lett. {\bf B417}, 173 (1998)
			and references therein.}

\bibitem{Latorre}    {J.I. Latorre, P. Pascual, and S. Narison,
			Z. Phys. {\bf C34}, 347 (1987).}

\bibitem{Wilson}   {K.G. Wilson,
                        Phys. Rev. {\bf D10}, 2445 (1974).}

\bibitem{K&S}       {M. Kremer and G. Schierholz,
			Phys. Lett. {\bf B194}, 283 (1987).}

\bibitem{Rothe}	   {H. J. Rothe, ``Lattice Gauge Theories''
			(World Scientific, 1997) 1.}

\bibitem{rabbit:3Q}    {T.T. Takahashi, H. Suganuma, Y. Nemoto 
			and H. Matsufuru,
			Phys. Rev. {\bf D65}, 114509 (2002).}

\bibitem{Banks}       {T. Banks and A. Casher,
                        Nucl. Phys. {\bf B169}, 103 (1980).}

\bibitem{Martinelli} {G. Martinelli, C. Pittori, C.T. Sachrajda,
			M. Testa, and A. Vladikas,
			Nucl. Phys. {\bf B445}, 81 (1995).}


\bibitem{Narison2}  { S. Narison and R. Tarrach,
			Phys. Lett. {\bf B125}, 217 (1983).}


\bibitem{Alad}      {K. Aladashvili and M. Margvelashvili,
                        Phys. Lett. {\bf B372}, 299 (1996).}

\bibitem{Beneke}      {M. Beneke and H.G. Dosch,
                        Phys. Lett. {\bf B284}, 116 (1992).}

\bibitem{Polyakov}    {M.V. Polyakov and C. Weiss,
			Phys. Lett. {\bf B387}, 841 (1996).}

\bibitem{DOIS:T}  {T. Doi, N. Ishii, M. Oka and H. Suganuma,
			Nucl. Phys. {\bf A} (2003) in press, hep-lat/0212006;
			Proc. Int. Conf. on
			``Quark Confinement and the Hadron Spectrum V'',
			Italy, 2002, hep-lat/0212025 .}


\end{references}
\end{document}